\documentclass[a4paper,11pt]{article}
\usepackage[english]{babel}
\usepackage[latin1]{inputenc}
\usepackage[T1]{fontenc} % Computer Modern (CM) fonts
\usepackage{ae,aecompl} % and: dvips -Pcmz -Pamz macros_aaa.dvi
\usepackage{amssymb}
\usepackage{amsmath}
\usepackage{graphicx}
\usepackage{subfig}
\usepackage{fancyhdr}
\IfFileExists{.pdfmode}{%
  
  \usepackage{ae,aecompl} % and: dvips -Pcmz -Pamz *.dvi
  }{
  
  }

\begin{document}

\thispagestyle{plain}
%Header of the paper
\begin{center}
\Huge{Modeling Trojan dynamics:}\\
\huge{diffusion mechanisms through resonances}
\end{center}

%Authors
\begin{center}
%First author data
\Large{Roc\'io Isabel P\'aez}\\
\small{Dip. di Matematica, Universit\`a di Roma ``Tor Vergata'', Italy}\\
\small{\textsc{paez@mat.uniroma2.it}}\\
\vspace{0.3cm}
%Second author data
\Large{Christos Efthymiopoulos}\\
\small{Research Center for Astronomy and Applied Mathematics, Academy of Athens, Greece}\\
\small{\textsc{cefthim@academyofathens.gr}}\\
%Copy&Paste if there are more.
\end{center}

%Abstract
\noindent
{\bf Abstract:} In the framework of the ERTBP, we study an example of
the influence of secondary resonances over the long term stability of
Trojan motions. By the integration of ensembles of orbits, we find
various types of chaotic diffusion, slow and fast. We show that the
distribution of escape times is bi-modular, corresponding to two
populations of short and long escape times. The objects with long
escape times produce a power-law tail in the distribution.
\begin{center}
\line(1,0){250}
\end{center}

\section{Resonances}
We study an example of mass parameter $\mu =0.0041$ and eccentricity
$e' = 0.02$ of the primary in the framework of the ERTBP.  Following
P\'{a}ez \& Efthymiopoulos 2014 (hereafter,
  P\&E14), we describe Trojan orbits in terms of
modified Delaunay variables given by

\begin{displaymath}
x = \sqrt{a} -1, \quad y = \sqrt{a} \left( \sqrt{1-e^2} -1 \right), \quad \Delta u = \lambda - \frac{\pi}{3} - u_0, \quad \omega~~,
\end{displaymath}
where $a$, $e$, $\lambda$ and $\omega$ are the major semi-axis,
eccentricity, mean longitude, and argument of the perihelion
of the Trojan body, and $u_0$ is such that $\Delta u = 0$ for
the $1$:$1$ short period orbit at $L_4$.

In this problem, the secondary resonances (see
P\&E14) are of the form $m_f\omega_f+m_s\omega_s+
m_g\omega_g=0$, involving the fast frequency $\omega_f$, the synodic
frequency $\omega_s$ and the secular frequency $\omega_g$ of the
Trojan body. Resonances are denoted below as [$m_f$:$m_s$:$m_g$]. The
most important resonances, called the 'main' secondary resonances,
correspond to the condition $\omega_f - n \omega_s = 0$
([$1$:$-n$:$0$]). For $\mu=0.0041$, this corresponds to
[$1$:$-6$:$0$].

\begin{figure}
\vspace*{-0.5 cm}
\hspace*{-0.5cm}
 \includegraphics[width=0.4\textwidth,angle=270]{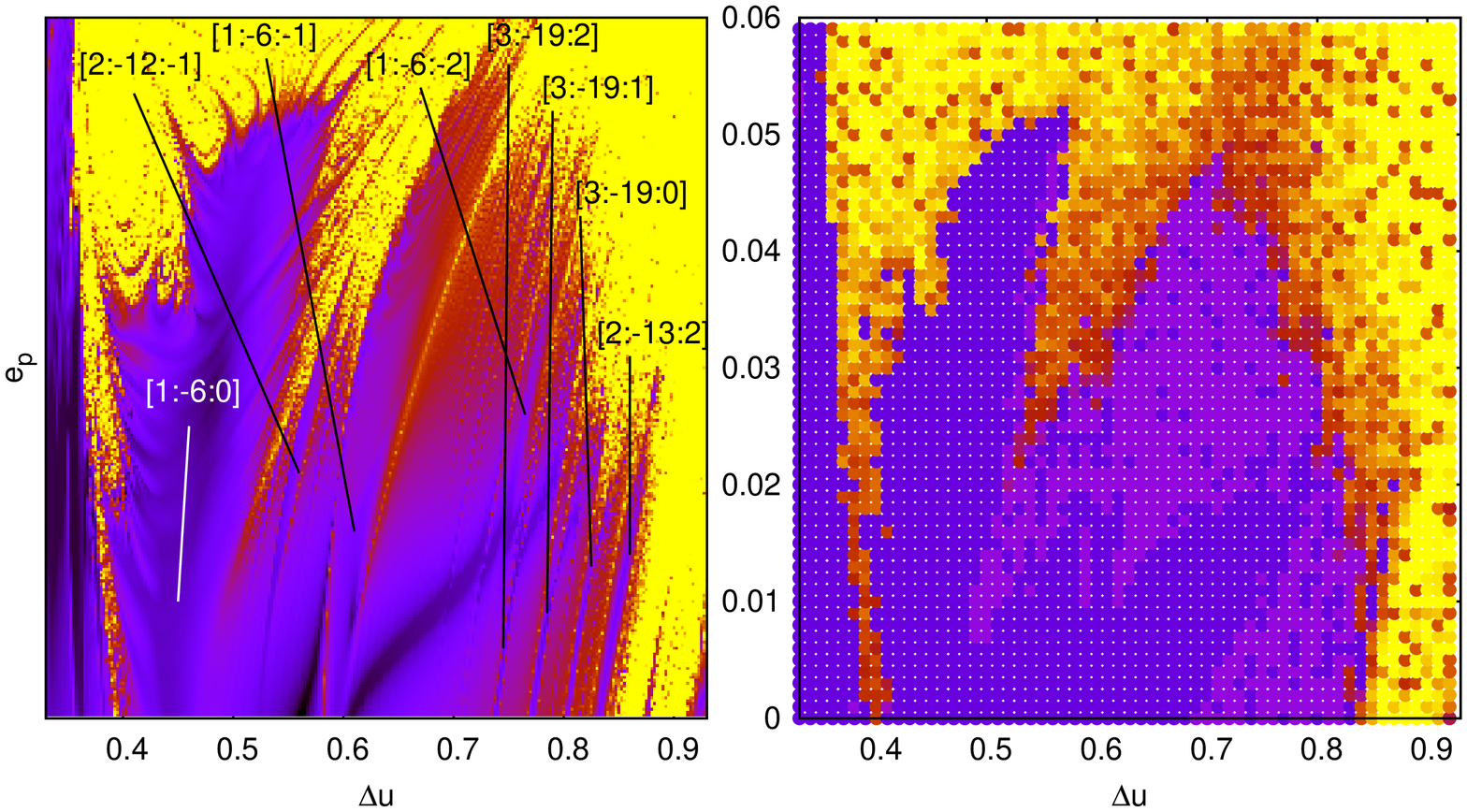} 
 \includegraphics[width=0.40\textwidth,angle=270]{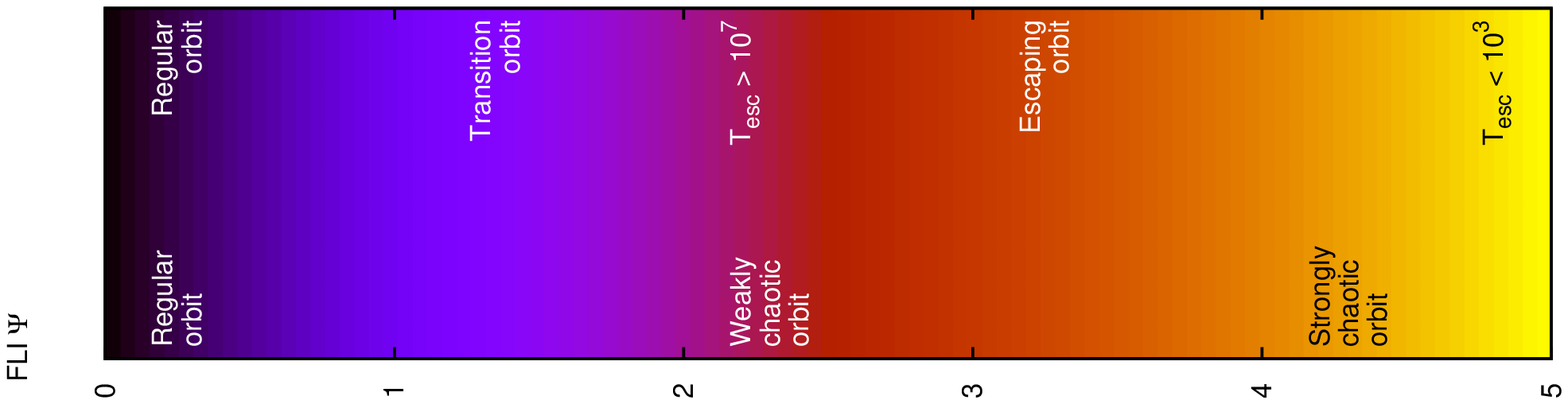}

\vspace*{-5.1cm} 
\hspace*{10.5cm}
 \includegraphics[width=0.25\textwidth,angle=270]{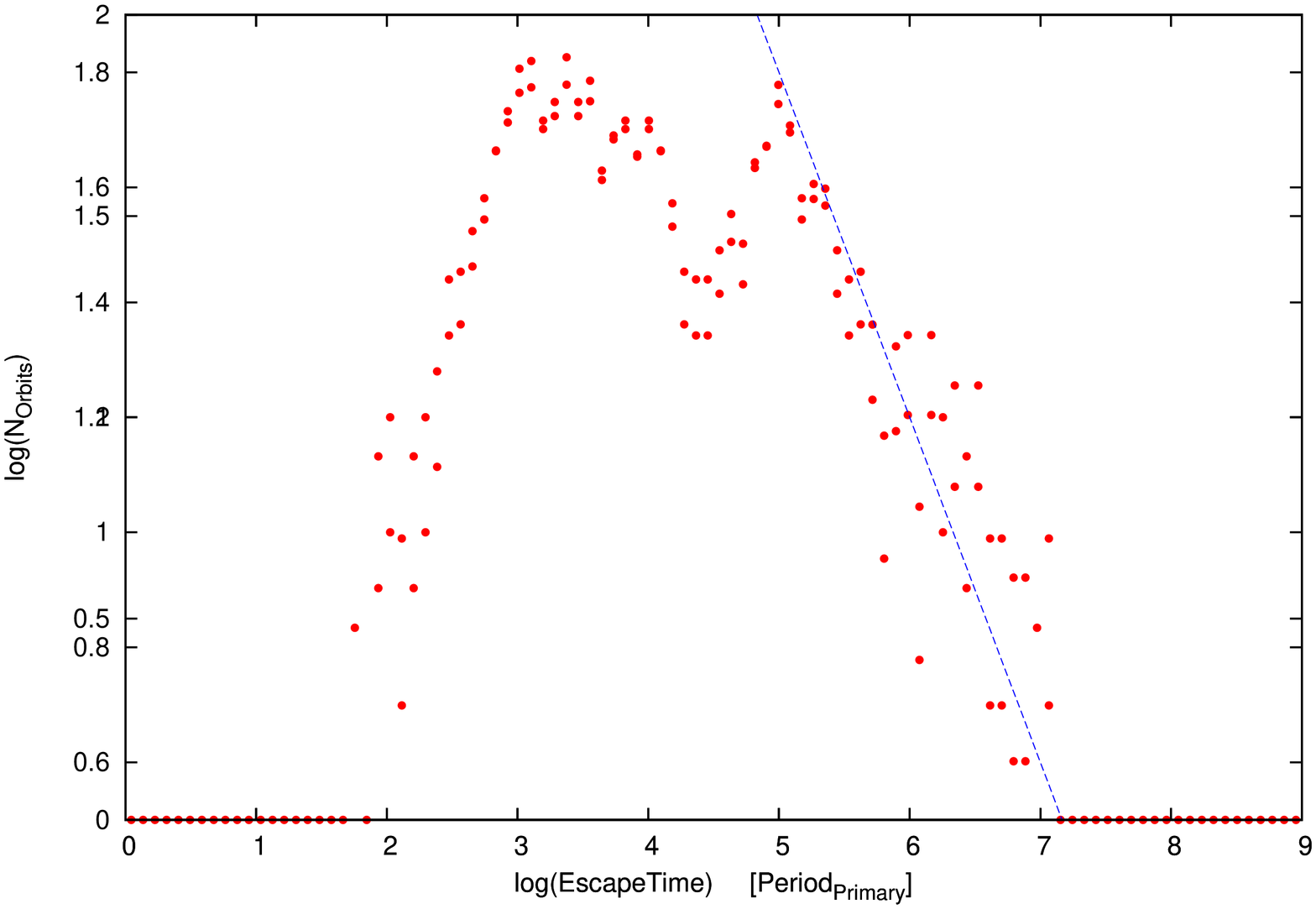} 
\vspace*{0.4 cm}
 \caption{Left: FLI map for initial conditions described in the text
   where various secondary resonances are distinguished in the space
   of proper elements ($\Delta u$,$e_p$). Middle: Color distribution
   of escaping times for the same initial conditions (color scale
   indicated). Right: distribution of the escaping times of the
   orbits.}
   \label{fig1}
\end{figure}

\vspace*{-0.5cm}
\section{Diffusion and stability}

Numerical experiments show that, for $e' > 0$, at least two different
mechanisms of diffusion are present.  Along non-overlapping
resonances, a slow (and practically undetectable) Arnold-like
diffusion (Arnold, 1964) takes place. On the other
hand, for initial conditions along partly overlapping resonances, due
to the phenomenon of pulsating separatrices
(P\&E14), we observe a faster 'modulational'
diffusion (Chirikov et al., 1985) leading to
relatively fast escapes.

In order to distinguish which parts of the resonant web provide each
behavior, we integrate 3600 initial conditions with $0.33 \leq \Delta
u \leq 0.93$ and $0 \leq e_p \leq 0.06$, where $\Delta u$ (libration
angle) and $e_p$ (proper eccentricity) are proper elements (see
Efthymiopoulos and P\'{a}ez, this volume). We visualize the resonance
web by color maps of the Fast Lyapunov Indicator FLI
(Froeschl\'{e} et al., 2000) of the orbits.
The resonances are identified by Frequency Analysis (Laskar,
  1990). We integrate all orbits up to $5$ different
integration times along $10^7$ periods of the primaries.  After each
integration, the initial conditions are categorized as {\bf
  \emph{Regular}} (if $\Psi (t) < \log_{10} (\frac{N}{10})$, where
$\Psi$ denotes the FLI value and $N$ is the total number of
integration periods), {\bf \emph{Escaping}} (if the orbit undergoes a
sudden jump in the numerical energy error greater than $10^{-3}$) or
{\bf \emph{Transition}} (non Regular nor Escaping).

{\small
\begin{center}
\vspace*{0.1cm}
\begin{tabular}{| c c c c |}
\hline
  N. of periods  & Regular Orb & Transition Orb & Escaping Orb \\
\hline
$10^3$ & 1220 ($33.8 \%$) & 2027 ($56.3 \%$) & 353 ($09.9\%$) \\
$10^4$ & 1263 ($35.0 \%$) & 1388 ($38.5 \%$) & 946 ($26.5\%$) \\ 
$10^5$ & 1296 ($36.0 \%$) & 966 ($26.8 \%$) & 1338 ($37.2\%$) \\
$10^6$ & 1299 ($36.1 \%$) & 699 ($19.4 \%$) & 1602 ($44.5\%$) \\
$10^7$ & 1309 ($36.3 \%$) & 603 ($16.8 \%$) & 1688 ($46.9\%$) \\ 
\hline
\end{tabular}
\vspace*{0.1cm}
\end{center}
\vspace{-0.1cm}
}

After $10^7$ periods, $46.9\%$ of the orbits have escaped.  However, a
significant portion ($16.8\%$) still remain trapped, despite having a
high FLI value.  Figure \ref{fig1} resumes the results.  The histogram
in the right panel shows two distinct timescales. The first peak
($10^3$ periods), corresponds to fast escapes, and the second ($10^5$
periods), to slow escapes.  When we compare the FLI map (left) with
the color distribution of the escaping times (middle), we find that
the majority of fast escaping orbits lay within the chaotic sea
surrounding the secondary resonances. The thin chaotic layers
delimiting the resonances provide both slowly escaping orbits and
\emph{transition} orbits (\emph{sticky} set of initial conditions that
do not escape after $10^7$ periods). For escaping orbits, beyond $t
\sim 10^5$ periods, the distribution of the escape times is given by
$P(t_{esc})\propto t_{esc}^{-\alpha},\,\alpha\approx 0.8$, while the
sticky orbits exhibit features of 'stable chaos' (Milani \&
  Nobili, 1992), since their Lyapunov times are much
shorter than $10^7$ periods.

\vspace{0.5cm}
%Acknowledgements
\noindent
\large{{\bf Acknowledgements:}} R.I.P. was supported by the Astronet-II Training
Network (PITN-GA-2011-289240).  C.E. was supported by the Research
Committee of the Academy of Athens (Grant 200/815).

\end{document}